\documentclass[aps,pra,10pt,singlecolumn,reprint,floatfix,superscriptaddress]{revtex4-1}
\usepackage[utf8]{inputenc} 
\usepackage{graphicx}
\usepackage{amsmath}
\usepackage{amsfonts}
\usepackage{mathtools}
\usepackage{float}
\usepackage{tabularx, booktabs}
\usepackage{color}
\usepackage{braket}
\usepackage{bbm}
\usepackage{comment}
\usepackage{hyperref}

\hypersetup{
	colorlinks=true,
	linkcolor=blue,
	citecolor=blue,
	filecolor=magenta,
	urlcolor=cyan,
	breaklinks=true
}
\newcommand{\papertitle}{Temporal Bell inequalities in non-relativistic many-body physics}
\newcommand{\ICFO}{ICFO-Institut de Ciencies Fotoniques, The Barcelona Institute of Science and Technology, Castelldefels (Barcelona) 08860, Spain.}
\newcommand{\ICREA}{ICREA, Pg. Lluís Companys 23, 08010 Barcelona, Spain.}

\DeclareSymbolFont{sfletters}{OML}{cmbrm}{m}{it}

\begin{document}
	
\title{\papertitle}
\author{Andrea Tononi}
\email{andrea.tononi@icfo.eu}
\affiliation{\ICFO}
\author{Maciej Lewenstein}
\affiliation{\ICFO}
\affiliation{\ICREA}


\begin{abstract}
Analyzing the spreading of information in many-body systems is crucial to understanding their quantum dynamics.
At the most fundamental level, this task is accomplished by Bell inequalities, whose violation by quantum mechanics implies that information cannot always be stored locally. 
While Bell-like inequalities, such as the one of Clauser and Horne, envisage a situation in which two parties perform measurements on systems at different positions, one could formulate temporal inequalities, in which the two parties measure at different times. 
However, for causally-connected measurement events, these extensions are compatible with a local description, so that no intrinsically-quantum information spreading is involved in such temporal correlations.
Here we show that a temporal Clauser-Horne inequality for two spins is violated for a nonzero time interval between the measurements if the two measured parties are connected by a spin chain. 
Since the chain constitutes the sole medium for the spreading of quantum information,  it prevents the immediate vanishing of Bell correlations after the first measurement and it induces violation revivals.
The dynamics we analyze shows that, as expected in a non-relativistic setup, the spreading of information is fundamentally limited by the Lieb-Robinson bound.
New insights on many-body quantum dynamics could emerge through future applications of our temporal Bell inequality to more general systems.
\end{abstract}

\maketitle 
\onecolumngrid

\section{Introduction}

Demonstrating that quantum mechanics is a complete theory has been a fundamental research topic with important applications in quantum information and communication. 
Inspired by the work of Einstein, Podolsky and Rosen \cite{einstein1935}, John Bell formulated his famous inequality \cite{bell1964}, which is violated by quantum mechanics. 
The experimental proofs of this violation \cite{friedman1972,aspect1982,pan1998} demonstrated that the quantum theory is incompatible with a local and deterministic viewpoint represented by hidden-variable theories.
In a similar fashion, Clauser–Horne–Shimony–Holt (CHSH) \cite{clauser1969} (and later Clauser and Horne (CH) \cite{clauser1974}) conceived other inequalities for binary-choice measurement correlators (or probabilities) on two separated parts of a quantum system.
A few years later, Leggett and Garg analyzed the case of repeated measurements in time on a system \cite{leggett1985}, demonstrating that also the time evolution and time correlations are intrinsically quantum and not classical. 
More recently, Fritz \cite{fritz2010} included the time variable in CHSH inequalities, while Kofler and Brukner explored the role of time and causality  \cite{kofler2008, brukner2014}.

In contemporary quantum foundations, Bell inequalities are defined in a device-independent framework that assumes $n$ spatially-separated parties $i=1,...,n$, each performing on the system one of the $m$ measurements $O_{1}^{(i)},...,O_{m}^{(i)}$, each one with $d$ possible outcomes $o_{j,1}^{(i)},...,o_{j,d}^{(i)}$.
All information is contained in the set of conditional probabilities $\{\mathcal{P}_{o_{j_1,l_1}^{(1)},...,o_{j_n,l_n}^{(n)}}(O_{j_1}^{(1)},...,O_{j_n}^{(n)})\}_{i, j_i,l_i}$ which, if the probabilities admit a local factorization, is constrained by Bell inequalities \cite{tura2014, tura2016}.
In this formalization, all parties are {implicitly} assumed to perform simultaneous instantaneous measurements. 
The aim of our paper is adapting this scenario to the context of time-delayed measurements, assuming in particular that information can only propagate non-relativistically within the system itself.

In the scenario of $n=2$ parties, as we show in the Methods, the conditional probabilities of \textit{causally-connected} measurement events admit a local description.
To preserve the quantum correlations in time, therefore, it is necessary to limit the transmission of information from a measuring party to the other.
For this scope, we note that, in the context of condensed matter systems, the Lieb-Robinson bound \cite{lieb1972} provides a finite propagation speed for the spreading of quantum information in a spin chain at the thermodynamic limit. 
We thus wonder how temporal Bell-type inequalities can be implemented in such a non-relativistic many-body setup so that, due to the propagation medium, the quantum correlations survive in time.
These inequalities, as a theoretical tool, would complement existing methods to characterize the dynamics of quantum correlations such as, for instance, the out-of-time correlators \cite{roberts2016, aleiner2016, lin2018, riddell2019, wysocki2024}, or other entanglement estimators \cite{amico2008, horodecki2008, laflorencie2016, bera2017, bou-comas2024, osterloh2002, verstraete2004, verstraete2004b, popp2005, bourenanne2004, brydges2019, frerot2023}, comparatively assessed in the Appendix.

Here we formulate a temporal Clauser-Horne inequality in terms of probabilities of measuring spin operators at different consecutive times, and we show analytically that it is violated at small finite times for Bell-correlated antipodal spins of an XX spin chain.
In particular, we first formulate a general CH inequality that describes the situation of two parties measuring two observables at different times. 
Then, we implement it in a specific many-body system by considering a setup made of a spin pair connected by an XX spin chain.
We show that the exact time evolution of a Bell-correlated pair violates the temporal CH inequality for small time between the measurements, and also at large revival times. 
These revivals, however, are suppressed by increasing the chain length. 
We observe that the persistence in time of quantum correlations in this non-relativistic setup is ensured by the Lieb-Robinson bound \cite{lieb1972}.
Since this limit encodes the specific many-body physics of the model, it is more relevant as a bound for the spreading of information than the speed of light, which reflects the generic assumption of causality. 

Our CH temporal inequality can be used, therefore, as a quantum-information inspired tool for characterizing the non-relativistic information dynamics of spin chains.
Note however that, due to this aim, our work differs from the past studies \cite{osterloh2002, verstraete2004, verstraete2004b, popp2005, bourenanne2004, brydges2019, frerot2023} that typically focused on entanglement characterization in the ground or low-energy many-body states (cf. Appendix).
In the future, we expect to witness more implementations of temporal Bell inequalities for characterizing the correlations dynamics of diverse many-body systems, including for instance Bose-Einstein condensates \cite{schmied2016, fadel2018, colciaghi2023}.

\section{Temporal Clauser-Horne inequality}
We consider two observers named Alice and Bob possessing different parts of a bipartite physical system described by the Hamiltonian $H$. At time $T=0$, Alice chooses randomly to measure either the observable $A_1$ or $A_2$, respectively obtaining either $a_1$ or $a_2$ as binary $\pm 1$-valued outcomes. 
Then, at time $T=t\geq 0$, Bob chooses randomly to measure either the observable $B_1$ or $B_2$, obtaining analogously a $\pm 1$-valued outcome $b_1$ or $b_2$. 
We denote with $p_{a_i b_j}(A_i,B_j(t))$ the conditional probability of observing the outcomes $a_i$ and $b_j$ given that $A_i$ and $B_j$ are measured. 
Note that, in this paper, the time evolved operators of Bob are calculated in the Heisenberg picture as $B_j(t) = e^{i H t} B_j e^{-i H t}$.

In 1974, Clauser and Horne (CH) formulated an inequality for the sum of various conditional probabilities in the form of $p_{a_i b_j}(A_i,B_j(0))$, and found that it is violated by quantum mechanics. Here we extend the CH inequality to the $t\geq 0$ case, deducing the following result (see Methods):
\begin{align}
\begin{split}
& 0 \leq I_{CH}(t) \leq 1,
\label{temporalCHineq}
\\
& I_{CH}(t) = p_{11}(A_1,B_2(t)) + p_{-1-1}(A_1,B_1(t)) ++p_{11}(A_2,B_1(t)) - p_{11}(A_2,B_2(t)).
\end{split}
\end{align}
This inequality is valid for generic Hermitian observables satisfying $A_i^2 = \mathbbm{1} = B_i^2$. 
It can be used directly, 
or easily generalized, to analyze the dynamics of quantum correlations in non-relativistic systems.

For concreteness, we will implement it
in this paper for spin-$1/2$ states $\ket{\uparrow}$, $\ket{\downarrow}$, and we will choose the spin observables $A_1 = \sigma^z$, $A_2 = \sigma^x$, $B_1 = (\sigma^z + \sigma^x )/\sqrt{2}$, and $B_2 = (\sigma^z - \sigma^x )/\sqrt{2}$, where $\sigma^x$, $\sigma^y$ and $\sigma^z$ are the Pauli matrices. 
For this choice, the Bell pair $(\ket{\uparrow\uparrow} + \ket{\downarrow\downarrow})/\sqrt{2}$ violates the temporal CH inequality maximally at $t=0$.
Indeed, we find for this state that $I_{CH}(0) = (1+\sqrt{2})/2 \approx 1.207 > 1$. 
Therefore, the temporal CH inequality Eq.~\eqref{temporalCHineq} is violated at $t=0$ by quantum mechanics, although it is respected by a local hidden-variable theory. 
Analyzing its eventual violation for $t > 0$ requires to specify the dynamics of the system.
We will describe the time evolution of $I_{CH}(t)$ in the following sections, by considering a many-body implementation in which Alice's and Bob's spins are located at the antipodal sites of a one-dimensional spin chain.

\section{Spin chain setup: initial state, Hamiltonian and time evolution}
We consider a chain of $N\geq 2$ spin-$1/2$ states with open boundary conditions. Alice and Bob measure, respectively, the spins located at the sites $1$ at $T=0$ and $N$ at $T=t$. 
The state at $T=0^-$ (before any of their measurements) is assumed to be the tensor product $\ket{\psi(0^-)} = \ket{\phi} \otimes \ket{\downarrow}_{2} \otimes \cdots \otimes \ket{\downarrow}_{N-1},$ where $\ket{\phi} = \left( \ket{\uparrow}_1 \otimes \ket{\uparrow}_N + \ket{\downarrow}_1 \otimes \ket{\downarrow}_N \right) / \sqrt{2}$ is a maximally-entangled Bell pair. 
When Alice measures the observable $A_i$ at $T=0$, the state is projected onto $\ket{\psi_{a_i}^{A_i}(0)} = \Pi_{a_i}^{A_i} |\psi(0^-) \rangle$, where $\Pi_{a_i}^{A_i} = (\mathbbm{1}+a_i A_i)/2$ is the projection operator into the subspace of $A_i$ corresponding to the outcome $a_i$. This is the initial state of the system. 

In the Heisenberg picture, the initial state $\ket{\psi_{a_i}^{A_i}(0)}$ does not evolve in time. We thus calculate the time-dependent conditional probabilities of Eq.~\eqref{temporalCHineq} as: 
\begin{align}
\label{condprobs}
p_{a_i b_j}(A_i,B_j(t)) = \bra{\psi_{a_i}^{A_i}(0)} \Pi_{b_j}^{B_j(t)}  \ket{\psi_{a_i}^{A_i}(0)},
\end{align}
where $\Pi_{b_j}^{B_j(t)} = (\mathbbm{1}+b_j B_j(t))/2$ is the projection operator corresponding to the outcome $b_j$ of the measurement $B_j(t)$. 
To explicitly calculate these quantities we need first to specify the Hamiltonian $H$ of the spin chain. 

We choose the XX Hamiltonian in transverse field with open boundary conditions:
\begin{equation}
H = -\frac{J}{2} \sum_{i=1}^{N-1} \left( \sigma^x_i \sigma^x_{i+1} + \sigma^y_i \sigma^y_{i+1} \right) - \frac{\mu}{2} \sum_{i=1}^N \left( \sigma^z_i + \mathbbm{1}_i \right), 
\label{XXHamiltonian}
\end{equation}
with $J$ and $\mu$ coupling constants of the model. 
To describe the time evolution more easily we carry out a Jordan-Wigner transformation which maps the spin states $\ket{\downarrow}$ and $\ket{\uparrow}$ into the fermionic occupation states $\ket{0}$ and $\ket{1}$ \cite{sachdev2011}. The operators $\sigma^{\pm}_i = (\sigma^x_i \pm i \sigma^y_i)/2$ are then mapped into fermionic creation and destruction operators $f_i^\dagger$ and $f_i$ (see Methods for details). 
The Hamiltonian becomes
\begin{align}
\begin{split}
H = -J \sum_{i=1}^{N-1} \left( f_i^\dagger f_{i+1} + f_{i+1}^\dagger f_i \right) - \mu \sum_{i=1}^N f_i^\dagger f_i 
= \sum_{m=1, \atop k=k_m}^N \epsilon_k c_k^{\dag} c_k,
\label{mappedHamiltonian}
\end{split}
\end{align}
with $c_k^{\dag},c_k$ the fermionic operators in the diagonal basis of the Hamiltonian, and where $\epsilon_k = - 2 J \lambda_k - \mu$ is the spectrum, written in terms of  $\lambda_k = \cos k$ (the lattice constant is taken equal to $1$) and $k_m = \pi m/(N+1)$, $m=1,2,...N$.

Note that the Hamiltonian $H$ fully determines the dynamics of the conditional probabilities $p_{a_ib_j}(A_i,B_j(t))$, since it allows to calculate $B_j(t)$. These spin observables of Bob can indeed be expressed via the Jordan-Wigner transformation in terms of fermionic operators $f_j(t)$ and $f_j^\dagger(t)$, whose dynamics is known analytically. In particular, we find that 
$f_j^\dagger(t) = \sum_{i=1}^N G_{ij}(t) f_i^\dagger$, where we define the propagator $G_{ij}(t) = \sum_{m=1, \atop k=k_m}^N u_{ik} u_{jk} e^{i \epsilon_k t}$, with $u_{jk} = (-1)^{j-1} U_{j-1}(\lambda_k)/[\sum_{l=1}^{N} U_{l-1}^2(\lambda_k)]^{1/2}$ normalized eigenfunctions expressed in terms of the Chebyshev polynomials of second kind $U_{j-1}(\lambda_k) = \sin(j k)/\sin(k)$.

\section{Violation of the temporal CH inequality}
\begin{figure}[hbtp]
\centering
\includegraphics[width=0.98\columnwidth]{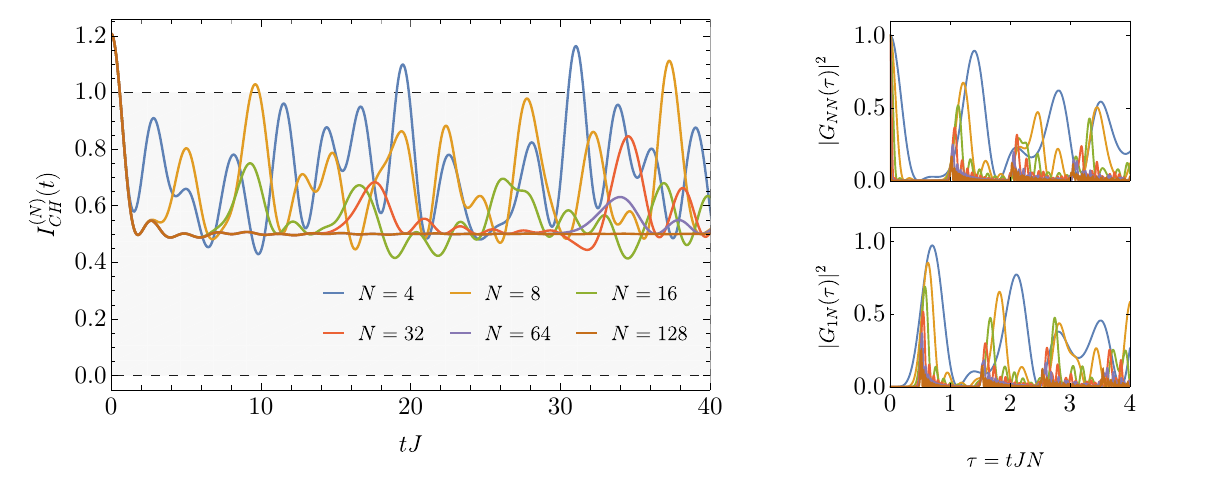}
\caption{Argument of the temporal Bell inequality $I_{CH}^{(N)}(t)$ versus $tJ$ for various lengths $N$ of the spin chain (left panel). The inequality $0 \leq I_{CH}^{(N)}(t) \leq 1$ (gray interval between dashed lines) is violated at small time and then restored by the time evolution, although there can be violation revivals which become less frequent as the number of spins $N$ increases. This is demonstrated by plotting the main contributions to the inequality $|G_{NN}(\tau)|^2$ and $|G_{1N}(\tau)|^2$ as a function of $\tau = tJN$ (right panels), which shows that these functions peak around multiples of, respectively, $tJ \sim N$ and $tJ \sim N/2$. At long times this behavior is blurred by the dephasing, which makes it less likely to have violations of the temporal inequality. The results shown in this figure correspond to $\mu/J=-1$.
}
\label{fig1}
\end{figure}

We denote the $N$-spins implementation of the temporal Bell inequality at Eq.~\eqref{temporalCHineq} by $I_{CH}^{(N)}(t)$, and we evaluate it analytically by calculating the time-dependent contractions $p_{a_i b_j}(A_i,B_j(t))$ of Eq.~\eqref{condprobs} under the Hamiltonian Eq.~\eqref{mappedHamiltonian} (see Methods). 
The argument of the temporal inequality $0 \leq I_{CH}^{(N)}(t) \leq 1$ can be expressed as
\begin{equation}
I_{CH}^{(N)}(t) = \frac{1}{2}  + \frac{\sqrt{2}}{4} \left(|G_{NN}(t)|^2 + |G_{1N}(t)|^2 + \text{Re}\big[ G_{NN}(t) \big]  \right),
\label{ICHexact}
\end{equation}
which is a known function of the parameters $N$, $tJ$, and $\mu/J$.
This formula for the $N$-sites spin chain is analytical and exact{, holding Eq.~\eqref{conjecture}.} 
At $t=0$ we find the expected result $I_{CH}^{(N)}(0) = (1+\sqrt{2})/2$ for any $N$, since $G_{NN}(0) = 1$ and $G_{1N}(0) = 0$. 
We then show the 
temporal behavior in Fig.~\ref{fig1} for a few values of $N$ and setting $\mu/J=-1$.

\vspace{1mm}


The temporal CH inequality is violated for a finite time interval $t$ between Alice and Bob measurements. 
Qualitatively, this means that quantum information cannot immediately spread within the chain and that the spins of the Bell pair $\ket{\phi}$ cannot {disentangle through their interaction with} the medium as soon as $t>0$ \cite{note}.
Since we cannot \textit{a priori} exclude Alice-Bob communication outside of the spin chain, the violation demonstrates the non-locality of quantum mechanics only for the time interval $t < t_c \propto N/c$ required by classical communication at the speed of light $c$ between the parties.
However, if, as stated since the beginning, we assume that information can only propagate within the spin chain, the inequality violations characterize mainly the non-relativistic quantum information dynamics.
In particular, the amount of quantum correlations that survive during the short-time evolution is quantified by $I_{CH}^{(N)}(t)$.

The violation of the inequality occurs until the finite time
$t^{*} = (4-2\sqrt{2})^{1/2}/(3 J^2+\mu^2)^{1/2}$, which we estimate for $N=3$ by expanding quadratically $I_{CH}^{(3)}(t)$ at small $t$ and solving $1=I_{CH}^{(3)}(t^{*})$. 
Actually, the estimate $t^{*}$ holds for any $N$ since the short-time behavior of $I_{CH}^{(N)}(t)$ is essentially $N$-independent and dominated by local processes occurring around the chain site $N$. 
Indeed, the main contributions to $I_{CH}^{(N)}(t)$ are provided by functions of the propagator  $G_{NN}(t) = \langle 0 | f_{N} f_{N}^\dagger(t) |0 \rangle$, which decays in a time scale $1/t^{*} \propto (3 J^2+\mu^2)^{1/2} \sim J$ since it represents the amplitude that a free fermion at site $N$ remains there until time $t$. 
Note that the dynamics and the consistency of the short-time violations depend on the values of $\mu/J$.
While for $|\mu/J| \lesssim 1$ the curves display the simple decay of Fig.~\ref{fig1}, for $|\mu/J| \gg 1$ we observe multiple short-time oscillations across $1$ of $I_{CH}^{(N)}(t)$, whose dynamics is still coincident with the one of $I_{CH}^{(3)}(t)$. 

The subsequent evolution of $I_{CH}^{(N)}(t)$ may again show breaking revivals, at which the temporal CH inequality is again invalidated since $I_{CH}^{(N)}(t) > 1$. 
The times at which these occur correspond roughly to multiples of $tJ \sim N$ and $tJ \sim N/2$, when, respectively, $G_{NN}(t)$ and $G_{1N}(t)$ peak (see the right panels of Fig.~\ref{fig1}). 
At these times the non-relativistic free fermionic excitations of Bob can reach the site of Alice, and the initial condition which violates the temporal inequality is approximately restored.
However, these breaking revivals are suppressed in the thermodynamic limit $N \gg 2$ irrespective of the $\mu/J$ value, and $I_{CH}^{(N)}(t)$ flattens around the value $1/2$ (see the $N=128$ case in Fig.~\ref{fig1}). 
Although qualitatively similar oscillations are also displayed by out-of-time correlators \cite{riddell2019}, the quantitative violations of the inequality are able to witness Alice and Bob's state entanglement and signaling occurring within the Hamiltonian dynamics \cite{mullerrigat2025}.

The dynamics we observe is ruled by the existence of the Lieb-Robinson bound  \cite{lieb1972}, according to which the spreading of quantum information in a spin-chain is bound to occur with a finite group velocity. 
An estimate of this quantity is provided by the excitations group velocity $v_g = (\partial \epsilon_k/\partial k) |_{k=\Bar{k}} = 2 J$, calculated at the momentum $\Bar{k} = \pi/2$ at which $\epsilon_{\Bar{k}} = E_{\psi(0^-)}$, with $ E_{\psi(0^-) } = \braket{\psi(0^-) | H | \psi(0^-)} = -\mu$ the initial energy. 
We argue that the Lieb-Robinson bound regulates the dynamics at any time.
Indeed $v_g \approx 1/t^*$, which if we interpret $I_{CH}^{(N)}(t)$ in fermionic language, means that the fermion initially at site $N$ cannot jump at neighboring sites as soon as $t>0$. 
At later times, analogously, the role of the Lieb-Robinson bound is reflected by revivals peaking around multiples of $tJ \sim N$ and $tJ \sim N/2$.

\section{Conclusion}
We formulated a temporal CH inequality for a non-relativistic many-body system, showing that it is violated for small finite time. 
This result demonstrates that the spin chain acts as a propagation medium by limiting the velocity at which information can travel and, as such, it allows the preservation of quantum correlations at short time. 
From a new perspective, therefore, we find that the maximum theoretical speed for the propagation of information is constituted
by the Lieb-Robinson bound for our specific non-relativistic model. 

Our work provides a concrete implementation of quantum information concepts in the context of non-relativistic many-body systems. 
On the application side, the framework developed in our paper constitutes a new method to characterize quantum entanglement and temporal correlations in spin chains, complementary to those of past works \cite{roberts2016, aleiner2016, lin2018, wysocki2024, amico2008, horodecki2008, laflorencie2016, bera2017, bou-comas2024, wysocki2024}.
Future theoretical and experimental studies may extend our approach to more than two measuring parties (or more than two time instances), or may study numerically the dynamics of the temporal CH inequality under more general spin Hamiltonians{, even in higher dimensions}.
Another interesting problem is determining which combination of spin Hamiltonians and initial states can display the inequality-breaking revivals.

Finally, we point out that our temporal Bell inequality is also applicable (or easily extendable) to other non-relativistic systems such as Bose-Einstein condensates \cite{schmied2016, fadel2018, colciaghi2023}, for which it can witness quantum correlations in time.
These generalizations could further elucidate which features of the many-body dynamics can be distinctively captured via temporal Bell inequalities.

\section{Methods}

\subsection{Hidden-variable description in the absence of a propagation medium}
We show here that, in the absence of a many-body medium connecting the measuring parties, there is a hidden-variable description compatible with: i.~probabilities of repeated measurements by the same party at different time, and ii.~probabilities of causally-connected measurements by two parties at different times. 
We remind that the dynamics is assumed here to be non-relativistic.
The following proofs, formulated for projective measurements, can be generalized to positive operator-valued measurements.

The statement i.~can be demonstrated for a system in the mixed state $\rho(0)$ at $T=0$.
The measuring protocol involves two steps. At $T=0$ Alice measures $A_1$ and obtains $a_1$ as outcome, thus projecting the state to $\ket{A_1,a_1}$. Then she measures $A_2$ at $T=t$ and obtains $a_2$ as outcome, projecting on $\ket{A_2,a_2}$. 
The probability of this process is 
\begin{align}
\mathcal{P}_{a_1 a_2}(A_1,A_2,t) = 
\braket{A_1,a_1 | \rho(0) | A_1,a_1} |\braket{A_2,a_2 | \mathcal{U}(t) | A_1,a_1}|^2,
\end{align}
where $\mathcal{U}(t)$ is the time-evolution operator of the system. 
This probability can also be expressed in terms of the hidden variables $\lambda=\{a_1',a_2'\}$, with probability distribution $\pi (\lambda) = \delta_{a_1 a_1'} \delta_{a_2 a_2'}$, as 
\begin{align}
\mathcal{P}_{a_1 a_2}(A_1,A_2,t) = 
\sum_{\lambda =\{a_1',a_2'\}} \pi (a_1',a_2') D_{a_1'}(A_1, \lambda) D_{a_1' a_2'}(A_1,A_2,t,\lambda),
\end{align}
where we introduce the deterministic probabilities \cite{bera2017} of being in the initial state $D_{a_1'}(A_1, \lambda) = \braket{A_1,a_1' | \rho(0) | A_1,a_1'} $ and of evolving from the initial to the final state $D_{a_1' a_2'}(A_1,A_2,t,\lambda) = |\braket{A_2,a_2' | \mathcal{U}(t) | A_1,a_1'}|^2$. 

Similarly, we can demonstrate the statement ii.~for a bipartite system in the state $\rho(0)$ at $T=0$. 
Alice measures $A$ at time $T=0$, obtains the outcome $a$ and projects the state on $\ket{A,a}$. The resulting state passes immediately to Bob, who measures $B$ at time $T=t$ and obtains $b$, thus projecting on $\ket{B,b}$. 
The probability of this process is
\begin{align}
\begin{split}
\mathcal{P}_{a b}(A,B,t) = 
\braket{B,b | \text{Tr}_A [\rho(A,a,t)] | B,b},
\end{split}
\end{align}
where $\rho(A,a,t) = \braket{A,a | \rho(0) | A,a} \mathcal{U}(t) \ket{A,a}\bra{A,a} \mathcal{U}^{\dagger}(t)$.
This probability can be rewritten with the hidden-variables $\lambda=\{a',b'\}$, with probability distribution $\pi (\lambda) = \braket{B,b' | \text{Tr}_A [\rho(A,a',t)] | B,b'}$, as 
\begin{align}
\begin{split}
\mathcal{P}_{a b}(A,B,t) = 
\sum_{\lambda =\{a',b'\}} \pi (\lambda) D_{ab}(B,t, \lambda),
\end{split}
\end{align}
where $D_{ab}(B,t, \lambda) = \delta_{a a'} \delta_{b b'}$. 
In this case, determinism results from Bob receiving the state plus the information of Alice measurement. 
If instead, as considered in our paper, this transfer is limited by the finite propagation velocity of a non-relativistic medium connecting the measuring parties, the above probabilities cannot be formulated within a local hidden-variable theory.
As a consequence, quantum correlations violate our non-relativistic implementation of the temporal Bell inequalities.

\subsection{Derivation of the temporal Clauser-Horne inequality}
We derive here the temporal Clauser-Horne inequality of Eq.~\eqref{temporalCHineq}. 
The inequality derived in 1974 by Clauser and Horne, which contains no time, reads \cite{clauser1974}
\begin{align}
&-1 \leq I_{CH}' \leq 0,
\\
\nonumber
&I_{CH}' = p_{11}(A_1,B_1) + p_{11}(A_1,B_2) +p_{11}(A_2,B_1) - p_{11}(A_2,B_2) - P_A(1|A_1) - P_B(1|B_1).
\end{align}
where $P_A(1|A_1)$ or $P_B(1|B_1)$ denote, respectively, the probabilities that Alice or Bob measure the observables $A_1$ or $B_1$ and obtain $1$ as outcomes. These can be expressed as $P_A(1|A_1) = \sum_{b_i = \pm 1} p_{1b_i}(A_1,B_i)$ and $P_B(1|B_1) = \sum_{a_i = \pm 1} p_{a_i1}(A_i,B_1)$, where $i=1$ or $i=2$. 
We reformulate the inequality by choosing $i=1$ in these relations, then we use the identity $- p_{11}(A_1,B_1) - p_{1-1}(A_1,B_1) - p_{-11}(A_1,B_1) = p_{-1-1}(A_1,B_1) -1$, and finally we include the time dependence in Bob's operators. 
As a result, we obtain an equivalent inequality for $I_{CH}(t) = I_{CH}'(t) + 1$, whose expression is given by Eq.~\eqref{temporalCHineq}.

\subsection{Jordan-Wigner transformation of the spin Hamiltonian and Heisenberg time evolution}
We discuss here the details of the Jordan-Wigner transformation \cite{sachdev2011}, which maps the spin Hamiltonian Eq.~\eqref{XXHamiltonian} to the fermionic diagonal Hamiltonian Eq.~\eqref{mappedHamiltonian}. 
The transformation maps the spin operators $\sigma^{\pm}_i$ to the fermionic operators  $f^{\dag}_i =  (\prod_{j<i} e^{-i\pi \sigma^{+}_j \sigma^{-}_j }) \sigma^{+}_i$ and $\quad f^{}_i = ( \prod_{j<i} e^{i\pi \sigma^{+}_j \sigma^{-}_j } ) \sigma^{-}_i$ which satisfy the anticommutation relations $\{f^{\dag}_i,f^{}_j  \} = \delta_{ij}$ and $\{f^{\dag}_i,f^{\dag}_j  \} = 0 = \{f_i,f^{}_j  \}$. 
The inverse transformation is given by $\sigma^{+}_i = \prod_{j<i} (1 - 2 f^{\dag}_j f^{}_j ) f^{\dag}_i$ and $\sigma^{-}_i = \prod_{j<i} (1 - 2 f^{\dag}_j f^{}_j ) f^{}_i$.

We substitute the spin operators in the Hamiltonian of Eq.~\eqref{XXHamiltonian} and we obtain the Hamiltonian at the middle term of Eq.~\eqref{mappedHamiltonian}. 
Then, to derive the Hamiltonian at the rightmost term, we decompose the fermionic operators as $f_j^\dagger = \sum_{m=1, \atop k=k_m}^N u_{jk} \, c_k^\dagger$ and $f_j = \sum_{m=1, \atop k=k_m}^N u_{jk} \, c_k$, where the definition of the normalized functions $u_{jk}$ and of the quantum numbers $k$ is provided in the main text. 

The time evolution of fermionic operators under the diagonal Hamiltonian is worked out in the Heisenberg picture. By solving the Heisenberg equation for $c_k^\dagger(t)$ one obtains $c_k^\dagger(t) = e^{i \epsilon_k t} c_k^\dagger$. Substituting $c_k^\dagger(t)$ into the decomposition of $f_j^\dagger(t)$, and re-substituting again the inverse decomposition of $c_k^\dagger$, we obtain the formula of the main text $f_j^\dagger(t) = \sum_{i=1}^N G_{ij}(t) f_i^\dagger$. This formula allows to calculate the time evolution of $f_j^\dagger(t)$ in terms of the propagator $G_{ij}(t)$, and therefore of all spin observables expressed in terms of fermionic operators. In particular, we will use the identity $\sigma_{N}^x(t) = \prod_{j<N} [1 - 2 f^{\dag}_j(t) f^{}_j(t) ] [f^{\dag}_N(t) + f^{}_N(t) ]$. 

\subsection{Derivation of Eq.~\eqref{ICHexact}}
The temporal Bell inequality of Eq.~\eqref{temporalCHineq} contains four conditional probabilities in the form of Eq.~\eqref{condprobs}. Here we evaluate them explicitly for the spin chain setup described in the main text. The procedure below will lead us to Eq.~\eqref{ICHexact}.

\vspace*{-3mm}
\subsubsection{Calculation of $p_{11}(A_1,B_2(t))$}
\vspace*{-3mm}
In the event described by this probability, Alice measures $A_1 = \sigma^z_1$ and obtains $a_1=+1$. Then, at time $t$, Bob measures $B_2(t) = (\sigma^z_{N}(t) - \sigma^x_{N}(t))/\sqrt{2}$ and obtains the eigenvalue $b_2 = +1$.  
The conditional probability of this {event} is given by $p_{11}(A_1,B_2(t)) = \bra{\psi_{1}^{A_1}(0)} \Pi_{1}^{B_2(t)}  \ket{\psi_{1}^{A_1}(0)}$.

To calculate the initial state $\ket{\psi_{1}^{A_1}(0)}$, we express the projector $\Pi_{1}^{A_1}$ for the measurement $A_1 = \sigma^z_1$ in terms of fermionic operators {and we apply it on} $\ket{\psi(0^-)}$, obtaining $\ket{\psi_{1}^{A_1}(0)} = \frac{1}{\sqrt{2}} f_1^\dagger f_{N}^\dagger |0 \rangle$. 
We calculate the desired conditional probability by contracting the projector $\Pi_{1}^{B_2(t)}$ over this initial state, obtaining
\begin{align}
p_{11}(A_1,B_2(t)) = \frac{2-\sqrt{2}}{8}  + \frac{\sqrt{2}}{4} \langle 0 | f_{N} f_1 f_{N}^\dagger(t) f_{N}(t)  f_1^\dagger f_{N}^\dagger |0 \rangle + \frac{\sqrt{2}}{8} \langle 0 | f_{N} f_1 \sigma^x_N(t)  f_1^\dagger f_{N}^\dagger |0 \rangle.
\end{align}
By expanding the fermionic operators in terms of the propagator, the first contraction equals equals $\langle 0 | f_{N} f_1 f_{N}^\dagger(t) f_{N}(t)  f_1^\dagger f_{N}^\dagger |0 \rangle =  |G_{NN}(t)|^2 + |G_{1N}(t)|^2$, while the second contraction is $0$.
Indeed, the multiplication of all the string operators contained inside $\sigma^x_N(t)$ generates addends with unequal numbers of creation and destruction operators, all of which have zero vacuum expectation value.
Therefore:
\begin{equation}
p_{11}(A_1,B_2(t)) =  \frac{2-\sqrt{2}}{8} + \frac{\sqrt{2}}{4} [|G_{NN}(t)|^2 + |G_{1N}(t)|^2].
\label{p11A1B2}
\end{equation}

\vspace*{-3mm}
\subsubsection{Calculation of $p_{-1-1}(A_1,B_1(t))$}
\vspace*{-3mm}
In the event described by this probability, Alice measures $A_1 = \sigma^z_1$ and obtains $a_1=-1$. At time $t$ Bob measures $B_1(t) = (\sigma^z_{N}(t) + \sigma^x_{N}(t))/\sqrt{2}$, obtaining the eigenvalue $b_1 = -1$. We thus need to calculate $p_{-1-1}(A_1,B_1(t)) = \bra{\psi_{-1}^{A_1}(0)} \Pi_{-1}^{B_1(t)}  \ket{\psi_{-1}^{A_1}(0)}$. 

Evaluating this contraction is simple, because Alice's measurement outcome produces the initial state $\ket{\psi_{-1}^{A_1}(0)} = (1/\sqrt{2}) \ket{0}$. Since the particle vacuum does not evolve in time, the conditional probability coincides with its $t=0$ value
\begin{equation}
p_{-1-1}(A_1,B_1(t)) = \frac{1}{2} \bra{0} \Pi_{-1}^{B_1(0)}  \ket{0} = \frac{2+\sqrt{2}}{8}.
\label{pm1m1A1B1}
\end{equation}

\vspace*{-3mm}
\subsubsection{Calculation of $p_{11}(A_2,B_1(t)) - p_{11}(A_2,B_2(t))$}
\vspace*{-3mm}
Given the definition of the projectors, and of $B_1$ and $B_2$, this probability difference reads
$p_{11}(A_2,B_1(t)) - p_{11}(A_2,B_2(t)) =  \bra{\psi_{1}^{A_2}(0)} [\sigma^x_{N}(t) / \sqrt{2}] \ket{\psi_{1}^{A_2}(0)}$. 

The initial state, obtained by applying the projector $\Pi_{1}^{A_2(t)}$ onto $\ket{\psi (0^{-})}$, is given by $\ket{\psi_{1}^{A_2}(0)} = \frac{\sqrt{2}}{4} (1 + f_1^\dagger + f_N^\dagger + f_1^\dagger f_N^\dagger) \ket{0}$. After substituting {it in} the expression above, we obtain 16 different contractions to evaluate. Half of them are zero, because they involve the vacuum expectation value of unequal numbers of creation and destruction operators, while the other half yields
\begin{align}
p_{11}(A_2,B_1(t)) - p_{11}(A_2,B_2(t)) = 
\frac{\sqrt{2}}{8} \text{Re}\big[\bra{0} (1 + f_N f_1) \sigma^x_{N}(t) (f_1^\dagger + f_N^\dagger) \ket{0} \big].
\end{align}

In the product above, two contractions can be immediately calculated:
$\bra{0} \sigma^x_{N}(t) f_N^\dagger \ket{0} = G_{NN}^{*}(t)$, $\bra{0} \sigma^x_{N}(t) f_1^\dagger \ket{0} = G_{1N}^{*}(t)$. 
Calculating {the other two involves complicated combinatorics generated by}
the string operators contained in the Jordan-Wigner transformed operators. 
However, we conjecture that
\begin{align}
\text{Re}\big[ 
\bra{0} f_N f_1 \sigma^x_{N}(t) f_1^\dagger \ket{0} 
+ \bra{0} f_N f_1 \sigma^x_{N}(t) f_N^\dagger \ket{0}  \big] = 
\text{Re}\big[G_{NN}(t) - G_{1N}(t) \big],
\label{conjecture}
\end{align}
and
we verified \cite{sneg} that the exact analytical form of the left- and right-hand sides coincides for $2\leq N \leq 5${, and with exact numerics} up to $N=8$ \cite{mullerrigat2025}. 
In the absence of a general proof, we assume the relation to hold for any value of $N$.
The desired probability sum is{, therefore} 
\begin{equation}
p_{11}(A_2,B_1(t)) - p_{11}(A_2,B_2(t)) =  \frac{\sqrt{2}}{4} \text{Re}\big[G_{NN}(t) \big].
\label{p11A2B1minusp11A2B2}
\end{equation}

In conclusion, we obtain the inequality of Eq.~\eqref{ICHexact} by summing the individual contributions calculated at the Eqs.~\eqref{p11A1B2}, \eqref{pm1m1A1B1}, and \eqref{p11A2B1minusp11A2B2}.

\section{APPENDIX: Entanglement characterization in many-body systems}
To further clarify our aim, here we comparatively review the main studies on the entanglement structure of many-body systems.

Initial studies on this topic \cite{osterloh2002} looked at the reduced density matrix for two spins of a spin chain, showing that even at criticality the entanglement vanishes for quite a short distance.
Later, findings of the paper series \cite{verstraete2004,verstraete2004b,popp2005} implied that, at criticality, when the correlation length diverges also does the entanglement length. 
With the advent of tensor networks methods, studies of entanglement in many-body systems mostly focused on ``area laws'' (for ground or low-energy states): in a bi-partitioned system into the blocks $A$ and $B$, the entropy $S(A)$ only depends on its boundary $\partial A$.
In one dimension, $S(A)$ is thus independent on the size of $A$ (except at criticality when it grows as $\log A$), while in higher dimensions $S(A)$ can be generally bounded by $|\partial A| \log(|\partial A|)$. 
Other methods of entanglement detection employed local measurements as entanglement witnesses \cite{bourenanne2004}, and efficient ways to use random unitaries for measuring low-order Renyi entropies were developed \cite{brydges2019}.
In our group, we have been focusing on the detection of entanglement and Bell correlations in many body systems by using second-order moments of local observables \cite{tura2014}, or non-complete data in a scalable way \cite{frerot2023}.

The above literaure focuses on entanglement characterization in the ground or low-energy states of many body systems.
Differently, here we monitor the out-of-equilibrium dynamics of a highly-excited state through a temporal Bell inequality.
The decay of the Bell inequality violation informs us on the propagation of information in the non-relativistic spin medium, providing an operative way to estimate the Lieb-Robinson bound.

\vspace{6mm}
\begin{acknowledgements}
We thank A. Ac\'in, H. Kurkjian, L. Piroli, M. Płodzień, L. Pricoupenko, and G. Müller-Rigat for useful discussions. 
A T acknowledges financial support of the Horizon Europe programme HORIZON-CL4-2022-QUANTUM-02-SGA via the project 101113690 (PASQuanS2.1).‌
ICFO-QOT group acknowledges support from:
European Research Council AdG NOQIA; 
MCIN/AEI (PGC2018-0910.13039/501100011033,  CEX2019-000910-S/10.13039/501100011033, Plan National FIDEUA PID2019-106901GB-I00, Plan National STAMEENA PID2022-139099NB-I00, project funded by MCIN/AEI/10.13039/501100011033 and by the “European Union NextGenerationEU/PRTR" (PRTR-C17.I1), FPI); QUANTERA MAQS PCI2019-111828-2;  QUANTERA DYNAMITE PCI2022-132919, QuantERA II Programme co-funded by European Union’s Horizon 2020 program under Grant Agreement No 101017733;
Ministry for Digital Transformation and of Civil Service of the Spanish Government through the QUANTUM ENIA project call - Quantum Spain project, and by the European Union through the Recovery, Transformation and Resilience Plan - NextGenerationEU within the framework of the Digital Spain 2026 Agenda;
Fundació Cellex;
Fundació Mir-Puig; 
Generalitat de Catalunya (European Social Fund FEDER and CERCA program, AGAUR Grant No. 2021 SGR 01452, QuantumCAT \ U16-011424, co-funded by ERDF Operational Program of Catalonia 2014-2020); 
Barcelona Supercomputing Center MareNostrum (FI-2023-3-0024); 
Funded by the European Union. Views and opinions expressed are however those of the author(s) only and do not necessarily reflect those of the European Union, European Commission, European Climate, Infrastructure and Environment Executive Agency (CINEA), or any other granting authority.  Neither the European Union nor any granting authority can be held responsible for them (HORIZON-CL4-2022-QUANTUM-02-SGA  PASQuanS2.1, 101113690, EU Horizon 2020 FET-OPEN OPTOlogic, Grant No 899794, QU-ATTO, 101168628),  EU Horizon Europe Program (This project has received funding from the European Union’s Horizon Europe research and innovation program under grant agreement No 101080086 NeQSTGrant Agreement 101080086 — NeQST); 
ICFO Internal “QuantumGaudi” project; 
European Union’s Horizon 2020 program under the Marie Sklodowska-Curie grant agreement No 847648;  
“La Caixa” Junior Leaders fellowships, La Caixa” Foundation (ID 100010434): CF/BQ/PR23/11980043.
\end{acknowledgements}

\end{document}